\documentclass[journal]{IEEEtran}

\ifCLASSINFOpdf

\else

\fi

\usepackage{lineno,hyperref}
\usepackage{amsmath}
\usepackage{amssymb}
\usepackage{graphicx}
\usepackage{subfigure}
\usepackage{threeparttable}
\usepackage{booktabs}
\usepackage{xcolor}
\newcommand{\rev}[1]{{\color{black} #1}}

\usepackage{todonotes}

\begin{document}

\bstctlcite{IEEEexample:BSTcontrol}

\title{Multiscale Wavelet Transfer Entropy with Application to Corticomuscular Coupling Analysis}
%
%
%

\author{Zhenghao~Guo,
        Verity~M.~McClelland,
        Osvaldo~Simeone,~\IEEEmembership{Fellow,~IEEE}, 
        Kerry~R.~Mills,
        and~Zoran~Cvetkovic,~\IEEEmembership{Senior~Member,~IEEE}
\thanks{This study was supported in part by the King’s-China Scholarship Council PhD Scholarship programme. The work of Osvaldo Simeone was supported by the European Research Council (ERC) under the European Union’s Horizon 2020 research and innovation programme (grant agreement No. 725731).}
\thanks{Zhenghao Guo, Osvaldo Simeone, and Zoran Cvetkovic are with Department of Engineering, King's College London, The strand, London, WC2B 4BG, UK.}
\thanks{Verity M. McClelland and Kerry R. Mills are with Department of Basic and Clinical Neuroscience, King's College London.}
\thanks{Copyright (c) 2021 IEEE. Personal use of this material is permitted. However, permission to use this material for any other purposes must be obtained from the IEEE by sending an email to pubs-permissions@ieee.org.}
}

\maketitle

\begin{abstract}
\textit{Objective:} Functional coupling between the motor cortex and muscle activity is commonly detected and quantified by cortico-muscular coherence (CMC) or Granger causality (GC) analysis, which are applicable only to linear couplings and are not sufficiently sensitive: some healthy subjects show no significant CMC and GC, and yet have good motor skills. The objective of this work is to develop measures of functional cortico-muscular coupling that have improved sensitivity and are capable of detecting \rev{both linear and non-linear interactions.} \textit{Methods:} \rev{A multiscale wavelet transfer entropy (TE) methodology is proposed. The methodology 
relies on a dyadic stationary wavelet transform  to decompose \rev{electroencephalogram (EEG) and electromyogram (EMG)} signals into functional bands of neural oscillations. Then, it applies TE analysis based on a range of embedding delay vectors to detect and quantify  intra- and cross-frequency band cortico-muscular coupling at different time scales.} \textit{Results:} Our experiments with neurophysiological signals substantiate the potential of the developed methodologies for detecting and quantifying information flow between \rev{EEG and EMG} signals for subjects with and without significant CMC or GC, including non-linear cross-frequency interactions, and interactions across different temporal scales. \rev{The obtained results are in agreement with the underlying sensorimotor neurophysiology.} \textit{Conclusion:} These findings suggest that the concept of multiscale wavelet TE provides a comprehensive framework for analyzing cortex-muscle interactions. \textit{Significance:} The proposed methodologies will enable developing novel insights into movement control and neurophysiological processes more generally.

\end{abstract}


%
\IEEEpeerreviewmaketitle

\section{Introduction}
\label{sec:introduction}
%
%
%
%
\IEEEPARstart{F}{unctional} coupling between the electroencephalogram (EEG) recorded over sensorimotor cortex and the electromyogram (EMG) of active muscles is usually quantified by using the spectral method of coherence estimation \cite{baker1997coherent}. \rev{Cortico-muscular coherence (CMC)} has been demonstrated in the $\beta$ band, $(16 - 32)$ Hz, in both humans and non-human primates \cite{baker1997coherent,graziadio2010developmental,mcclelland2012modulation}. Many authors have found \rev{CMC} in the $\beta$ band to be task dependent \cite{baker1997coherent}, related to motor performance \cite{graziadio2010developmental}, and modulated by different types and intensities of afferent stimuli \cite{mcclelland2012modulation}. These studies, and many others,   show that synchronous EEG-EMG recordings contain substantial information on neurophyisological mechanisms of movement control, and that some of the cortex-muscle interaction is linear.  However, details of the generation, propagation, and function of CMC are still unclear \cite{witham2010corticomuscular}. Towards exploring its origin and function, some studies have applied \rev{Granger causality (GC) or directed coherence (DC)} \cite{granger1969investigating,witham2010corticomuscular,witham2011contributions}. However, both CMC and GC analysis currently used in motor neurophysiology have limitations that impede their clinical use. In particular, conventional CMC and GC only capture linear associations between processes and are not sufficiently sensitive: for healthy subjects with good motor skills, the values of CMC and GC may even fall below the significance threshold \cite{mcclelland2012modulation,salenius2003synchronous}. Recently, the concept of cross-spectral coherence has been developed, capable of assessing synchrony  between distinct frequencies that arises as a result of non-linear coupling \cite{yang2016generalized,yang2016nonlinear,sinha2020cross}. However, its discrete frequency nature necessitates some prior knowledge about synchronised frequencies, or about the type of non-linearity and the direction of information propagation. The methodology is therefore most effective in combination with methods that analyse coupling across frequency bands, including the method proposed here.


Information theory provides principled tools to quantify the transmission of information, such as mutual information (MI), directed information (DI), and transfer entropy (TE). Their model-free nature and ability to capture both linear and non-linear relationships has spurred considerable interest in their potential for detecting and quantifying functional neurophysiological coupling \cite{lindner2011trentool,vicente2011transfer, lizier2014jidt,malladi2018mutual,jin2010linear,malladi2016identifying,yang2012new,gao2018electroencephalogram,zhou2018optimization,shi2018cross,so2016cross,zhao2018multiscale,chen2018multiscale}. 
Neurophysiological information transfer takes place within and across distinct functional bands
of neural oscillations. With the goal of unravelling principles and protocols of  neural communications we propose the concept of TE in the domain of a dyadic wavelet transform that approximately performs signal decomposition into these functional bands. For the analysis of \rev{cross-frequency coupling (CFC)} of neural oscillations using information theoretic measures two methodologies have recently been developed: Malladi {\sl et al.} proposed using MI in the domain of the \rev{discrete Fourier transform (DFT)} computed over adjacent non-overlapping signal segments \cite{malladi2018mutual}, whereas  Shi {\sl et al.} proposed TE between amplitudes/phases or overall intrinsic mode components obtained via empirical mode decomposition (EMD) \cite{shi2018cross}. The methodology proposed here is an extension of the MI approach in \cite{malladi2018mutual} that allows for the detection of the direction of information propagation, owing to its reliance on TE. Moreover, it performs  the analysis in the functional frequency bands of neural oscillations as advocated in \cite{malladi2018mutual}, which are physiologically more relevant than the interactions across instantaneous frequencies \cite{brittain2014oscillations}  considered in \cite{shi2018cross}. 

Estimation of TE requires representing the considered processes by {\sl embedding delay vectors} \rev{that consist of their samples separated in time by {\sl embedding delays}}  \cite{vicente2011transfer}. Adequate selection of embedding delays and the dimensions of embedding vectors, referred to as the {\sl embedding dimensions}, requires a careful balance between conflicting requirements: one must ensure that embedding vectors describe the considered processes well, whilst keeping their dimension small enough to allow  reliable TE estimation from limited data. To that end, various criteria have been developed \cite{yang2012new,cao1997practical,ragwitz2002markov,pan2020structure}. The two most commonly used are the practical and partly heuristic criterion by Cao \cite{cao1997practical} and  the  optimal search-based criterion by Ragwitz and Kantz \cite{ragwitz2002markov}, along with its efficient implementation \cite{zhou2018optimization}.  The selection of embedding vectors, however, depends on various assumptions about the underlying system, and remains not well understood in general \cite{pan2020structure}. Hence, in the applications of information theoretic measures in neurophysiology the embedding parameters are often set heuristically, based on domain specific knowledge, or governed by other methodological choices \cite{malladi2018mutual,jin2010linear,malladi2016identifying,gao2018electroencephalogram}.  

As signals in frequency bands of neural oscillations reflect mixtures of concurrent brain processes, one set of embedding vectors seems insufficient to describe complex brain processes. This is true even within a single function, where we typically observe phase-to-phase, phase-to-amplitude, amplitude to amplitude and more complex interactions \cite{shi2018cross}. Therefore, we propose TE analysis in the dyadic wavelet transform domain using embedding vectors that represent sub-band signals at different time scales \rev{and refer to this approach as {\sl multiscale wavelet transfer entropy.} In experiments with simulated data, the multiscale wavelet TE reliably and robustly detects information transfer between  processes.
Our empirical results obtained by applying this methodology to synchronous EEG-EMG  signals show that it detects cortico-muscular interactions regardless of whether subjects exhibit significant CMC or GC, with largely consistent intra- and cross-frequency band patterns
that are in agreement with the underlying motor neurophysiology.}

\rev{
The paper is organised as follows. Section \ref{sec:background} reviews relevant aspects of 
TE analysis. Section \ref{sec:methods} describes the proposed {\sl multiscale wavelet transfer entropy} methodology. Section \ref{sec:simulation} and Section \ref{sec:results} report the experimental results using simulated data and physiological recordings.
A discussion and conclusions are presented in Section \ref{sec:discussion} and Section \ref{sec:conclusion}.
}

\section{Background}
\label{sec:background}
\subsection{TE and its Estimation}
\label{sec:te}
Given two time series of interest \(\{x_t\}\) and \(\{y_t\}\), TE measures the amount of information that the past samples of the source process \(\mathbf{x}_{t-1}^{m}=(x_{t-m},x_{t-m+1},...,x_{t-1})\) provide about the target process' next state \(y_{t}\) given its past samples \(\mathbf{y}_{t-1}^{n}=(y_{t-n},y_{t-n+1},...,y_{t-1})\)  \cite{schreiber2000measuring}. 
TE is formally defined as
\begin{equation}
\begin{split}
\mathrm{TE}_{x \rightarrow y}(m, n)
&=\mathrm{I}(\mathbf{x}_{t-1}^{m} ; y_{t} | \mathbf{y}_{t-1}^{n})\\
&=\mathrm{E}\left[\mathrm{log}\left({\frac{p(y_{t}|\mathbf{x}_{t-1}^{m},\mathbf{y}_{t-1}^{n})}{p(y_{t}|\mathbf{y}_{t-1}^{n})}}\right)\right],
\label{eq:te}
\end{split}
\end{equation}
where \(\mathrm{I}(x;y|z)\) represents the conditional \rev{MI} of jointly distributed random variables \((x,y,z)\), and the expectation is taken over their joint probability distribution. The definition of the TE can be modified to account for an arbitrary source-target delay \(u \geq 0\) as \cite{lindner2011trentool}
\begin{equation}
\begin{split}
\mathrm{TE}_{x \rightarrow y}(m, n, u)
&=\mathrm{I}(\mathbf{x}_{t-1}^{m} ; y_{t+u} | \mathbf{y}_{t+u-1}^{n}),
\label{eq:tearbitrarydelay}
\end{split}
\end{equation}
where we have defined \(\mathbf{y}_{t+u-1}^{n}=(y_{t+u-n},...,y_{t+u-1})\).

The definition (\ref{eq:tearbitrarydelay}) implicitly assumes that processes \(\{x_{t}\}\) and \(\{y_{t}\}\) are jointly stationary Markov processes with parameters \(m\) and \(n\). When this is not the case, sensible causality measures should account for properly defined variables that capture the memory of observed processes \cite{vicente2011transfer}. 
We thus consider a form of (\ref{eq:tearbitrarydelay}) where state variables are defined from non-successive samples of the processes \cite{vicente2011transfer}. \rev{To this end, we define the delay embedding vectors}
\begin{subequations}
\begin{equation}
\mathbf{x}_{t-1}^{\tau_x,d_x}=(x_{t-(d_{x}-1) \tau_{x}-1},x_{t-(d_{x}-2) \tau_{x}-1}, \dots, x_{t-1}),\\
\end{equation}
\begin{equation}
\mathbf{y}_{t+u-1}^{\tau_y,d_y}=(y_{t-(d_{y}-1) \tau_{y}+u-1},y_{t-(d_{y}-2) \tau_{y}+u-1}, \dots, y_{t+u-1}),
\end{equation}
\label{eq:delayembeddingvector}%
\end{subequations}
\rev{where} \(\tau_{x}\) and \(\tau_{y}\) are known as embedding delays, and \(d_{x}\) and \(d_{y}\) are known as embedding dimensions for the time series \(x_t\) and \(y_t\), respectively. The TE is then redefined as
\begin{equation}
\begin{split}
\mathrm{TE}_{x \rightarrow y}(\tau_x, d_{x}, \tau_y, d_{y}, u)
=\mathrm{I}(\mathbf{x}_{t-1}^{\tau_x,d_{x}} ; y_{t+u} | \mathbf{y}_{t+u-1}^{\tau_y,d_{y}}) .
\label{eq:tearbitrarydelayembedding}
\end{split}
\end{equation}

Estimating the TE (\ref{eq:tearbitrarydelayembedding}) from a finite number of samples of the time series of interest is a complex task, which should account for the type of data and its properties \cite{lizier2014jidt}. We focus here on continuous-valued data, as is typical in neurophysiology. To simplify the notation in (\ref{eq:tearbitrarydelayembedding}), we first let \(\mathbf{x}^{-} \triangleq \mathbf{x}_{t-1}^{\tau_x,d_{x}}\), \(\mathbf{y}^{-} \triangleq \mathbf{y}_{t+u-1}^{\tau_y,d_{y}}\), and \(y \triangleq y_{t+u}\). Using this notation, the TE (\ref{eq:tearbitrarydelayembedding}) can be rewritten as sum of four differential entropies as \cite{gomez2015assessing}
\begin{equation}
\begin{split}
\mathrm{TE}_{x \rightarrow y}
&=\mathrm{I}(\mathbf{x}_{t-1}^{\tau_x,d_{x}} ; y_{t+u} | \mathbf{y}_{t+u-1}^{\tau_y,d_{y}}) \\ 
&=\mathrm{H}(y, \mathbf{y}^{-})-\mathrm{H}(\mathbf{y}^{-})-\mathrm{H}(y, \mathbf{y}^{-}, \mathbf{x}^{-})+\mathrm{H}(\mathbf{y}^{-}, \mathbf{x}^{-}).
\label{eq:tesimplied}
\end{split}
\end{equation}

\noindent
The TE in (\ref{eq:tesimplied}) can then be estimated by combining estimates of the individual  differential entropy terms, which can be obtained using either parametric or non-parametric methods \cite{beirlant1997nonparametric}. For neurophysiological data, non-parametric techniques, such as kernel and nearest-neighbor estimators have been shown to be most data-efficient and accurate \cite{lindner2011trentool,lizier2014jidt}. In the present study we use the Kraskov, St{\"o}gbauer, Grassberger (KSG) technique \cite{kraskov2004estimating} for TE estimation, which builds on the Kozachenko and Leonenko (KL) estimator \cite{Kozachenko1987Sample} of log-probabilities via nearest-neighbour counting \cite{lizier2014jidt}.

\subsection{Statistical Testing}
\label{sec:significance}
The TE between two independent time series is equal to 0. However, when it is estimated using a finite amount of data, this may not be the case. A key question is then whether a given estimate is statistically different from 0, and therefore provides sufficient evidence for interactions between observed processes. \rev{Towards establishing statistically significant interactions, all considered methods were applied to a 
total of 1000 pairs of independent random Gaussian sequences with the variances equal to the variances  of investigated EEG and EMG signals, and the obtained results were used to form empirical distributions of CMC, GC and TE that correspond to such independent processes. The  95\% confidence levels (CL) above which CMC, GC, and TE are considered to indicate genuine interactions are then set as 95 percentiles of the respective empirical distributions}. \rev{In the case of CMC, this level admits the closed form expression given in \cite{rosenberg1989fourier}}.

\section{Multiscale  Wavelet  Transfer Entropy}
\label{sec:methods}
\subsection{Dyadic Stationary Wavelet Transform}
\label{sec:swt}
Towards developing an understanding of information transfer within and across functional frequency bands of neural oscillations,  $\delta$  $(0-4)$ Hz, $\theta$  $(4-8)$ Hz, $\alpha$  $(8-16)$ Hz, $\beta$ $(16-32)$ Hz, and  $\gamma$ $(32-150)$ Hz bands,  we propose TE analysis in the domain of a wavelet transform that decomposes signals into components that approximately occupy these bands. In particular we propose using the discrete dyadic stationary wavelet transform (SWT) as detailed in the following.
\subsubsection{Dyadic stationary wavelet transform}
\begin{figure}[htbp]
    \centering
    \subfigure[SWT filter bank.]{
        \includegraphics[width=0.46\textwidth]{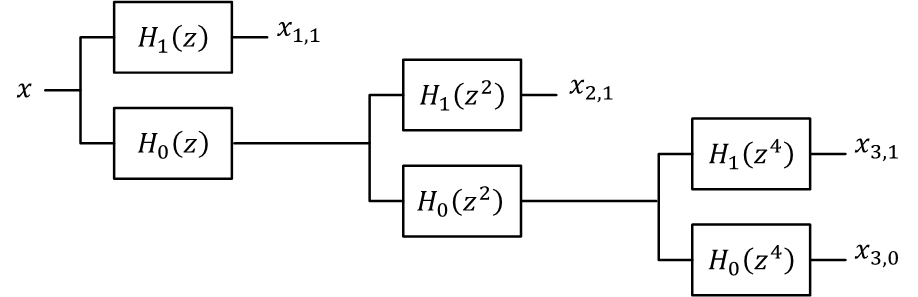}
        \label{figure:swtfb}
    }
    \subfigure[DWT filter bank.]{
        \includegraphics[width=0.46\textwidth]{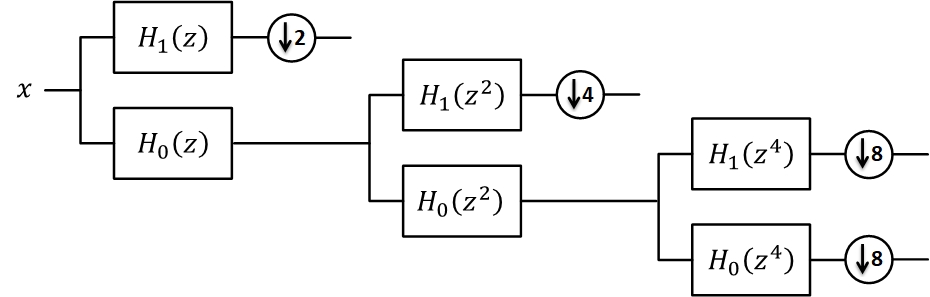}
        \label{figure:dwtfb}
    }
    \caption{\rev{The relationship between the stationary wavelet transform (SWT) and the conventional discrete-time wavelet (DTW) transform illustrated via  filter banks which implement them.The DWT involves a dyadic downsampling in the filter-bank channels, whereas the SWT does not perform the downsampling operation.}}
\end{figure}

The dyadic \rev{SWT} proposed here 
decomposes a signal $x$ into its $J+1$ subband components $x_{J,0},x_{j,1},j=1,\ldots,J$.  These are obtained by convolving $x$ using filters $h_{J,0},~h_{j,1},j=0,\ldots,J$, respectively, as
\begin{subequations}
\begin{equation}
x_{J,0}=h_{J,0} \ast x,
\end{equation}
and
\begin{equation}
x_{j,1}=h_{j,1} \ast x,~j=1,\ldots,J,
\end{equation}
\label{eq:swt}%
\end{subequations}
\rev{where} $J$ denotes the total number of {\sl scales of the wavelet transform indexed by $j$.} The filters are derived from two prototype filters $h_0$ and $h_1$ through an iterative procedure that is specified in the $z$-transform domain as
\begin{subequations}
\begin{equation}
H_{J,0}(z)=\prod_{k=0}^{J-1}H_0(z^{2^k}),
\end{equation}
\begin{equation}
H_{j,1}(z)=H_1(z^{2^{j-1}})\prod_{k=0}^{j-2}H_0(z^{2^k})~.
\end{equation}
\label{eq:filters}%
\end{subequations}
\rev{The} SWT, also known as the non-decimated wavelet transform, is a translation-invariant modification of the conventional discrete wavelet transform (DWT), that is obtained by removing the downsampling operators from the channels of the DWT, as illustrated in Figure \ref{figure:swtfb} and \ref{figure:dwtfb}. In addition to making the SWT be translation invariant, the absence of the downsampling operators in the SWT has other important benefits in this context: it retains considerably more data for TE estimation compared to the conventional DWT \rev{and it allows for a more flexible filter design under requirements of no information loss \cite{cvetkovic1998oversampled}.}

\subsubsection{Time-frequency resolution}
Filters $h_0$ and $h_1$ are  designed to be half-band low-pass and half-band high-pass, respectively. Hence, components $x_{j,1},~j=1,\ldots,J$ and $x_{J,0}$  occupy approximately frequency bands
$$\left({\frac{\pi}{2^j}},{\frac{\pi}{2^{j-1}}}\right), j=1,\ldots J, ~{\rm and}~ \left(0,{\frac{\pi}{2^J}}\right),$$
respectively. If signals are acquired using $1024$ Hz sampling, then $x_{3,1}$, $x_{4,1}$, $x_{5,1}$, $x_{6,1}$, $x_{7,1}$ and $x_{7,0}$ approximately represent
high $\gamma$, low $\gamma$, $\beta$, $\alpha$, $\theta$ and $\delta$ oscillations, respectively. Analogous correspondence of the frequency bands holds for other sampling frequencies that are powers of $2$ larger or equal than $256$ Hz, whilst for other sampling frequencies such a correspondence can be achieved by employing a re-sampling pre-processing step. 

The time resolution is governed by the filter length, and, as is inherent to all types of the wavelet transform, it is progressively higher with ascending frequencies. Thus, it is well suited for representing both transient events, which typically reside at high frequencies, as well as large scale processes composed of low-frequency oscillations. Additional flexibility in terms of the analysis at different time scales in the context of the TE analysis is achieved by representing the subband signals using different embedding parameters, as discussed in Section \ref{sec:multiscale}.

\subsubsection{Choice of filters}
The SWT is used in this work as a pre-processing step towards studying information flow in functional bands of brain waves. However, although reconstruction of signals from their SWT will not be considered, it is important to use filters that allow for a stable perfect reconstruction so that no information is lost in  the process of the subband decomposition. Moreover, it is beneficial to ensure that all the information is equally represented in the SWT domain, which is achieved when orthogonal two-channel filter banks \cite{vetterli}, such as those proposed by Daubechies \cite{daubechies}, or those implementing tight frames are used \cite{cvetkovic1998oversampled}, with the latter allowing for more flexible designs.


\subsection{Multiscale Wavelet Transfer Entropy}
\label{sec:multiscale}
Setting appropriate embedding dimension \(d\) and embedding delay \(\tau\) for the computation of the TE is a subtle issue. Towards motivating our concept of multiscale wavelet TE, we assess first the two most commonly used approaches \cite{lindner2011trentool,vicente2011transfer} proposed  by Ragwitz and Kantz  
\cite{ragwitz2002markov} and Cao \cite{cao1997practical}, using EEG and EMG signals collected synchronously during a motor control task (see Section \ref{sec:data}) \cite{mcclelland2012modulation}. For the clarity of the presentation, the SWT used in this assessment was performed to the depth that extracts the $\alpha$, $\beta$ and low $\gamma$ bands, but does not split the $(0-8)$ Hz band into $\delta$ and $\theta$ subbands.

\begin{figure}[htbp]
    \centering
    \includegraphics[width=0.48\textwidth]{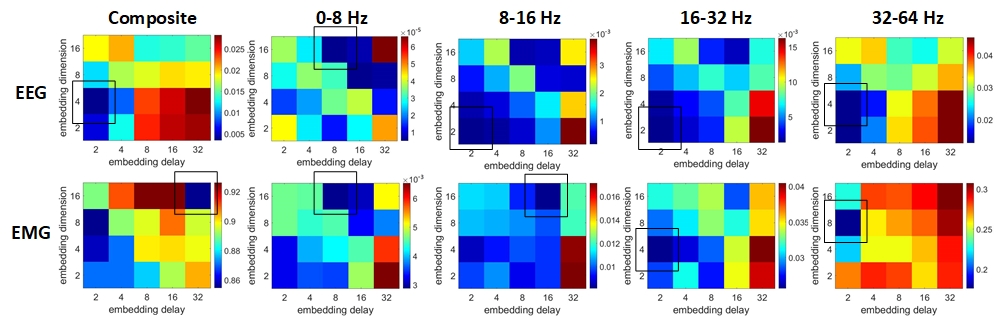}
    \caption{Ragwitz' criterion. \rev{Mean square prediction errors} of composite and subband EEG and EMG signals were ploted on the (\(d\)-\(\tau\)) plane with logarithmic scale. The black rectangle indicates the  \(d\)-\(\tau\) pairs \rev{that minimise the prediction error.}}
    \label{figure:ragwitz}
\end{figure}

The Ragwitz’s criterion estimates the embedding delays \(\tau\) and embedding dimensions \(d\) by searching in the (\(d\)-\(\tau\)) plane to identify the point  that minimizes the mean squared prediction error (MSPE)  of the process using  its embedding vector \cite{lindner2011trentool,ragwitz2002markov}. Figure \ref{figure:ragwitz} plots the prediction errors landscape of Ragwitz' criterion for the composite and subband EEG and EMG signals of Subject A. The obtained optimal embedding parameters appear counter-intuitive and inconsistent. For instance, for the composite EEG signal, optimal $\tau$ and $d$ are $2$ ($\sim 2$ ms) and $4$, respectively, suggesting prediction using only $6$ ms of  context,  whereas the optimal prediction of the subband component occupying the $(0-8)$ Hz band is achieved by  $\tau$ and $d$ equal to $8$ ($\sim 8$ ms) and $16$, respectively, which corresponds to a much larger context of approximately $120$ ms. The obtained embedding parameters in the $\alpha$ and $\beta$ bands are $\tau=2$ and $d=2$, which appears too low, both in terms of the dimension and the overall context of $\sim 2$ ms. The optimal embedding parameters for the composite EMG signal and $\tau$ and $d$ equal to $32$  ($\sim 32$ ms) and $16$, respectively, which corresponds to the context of around $480$ ms, whereas all subband components appear to be optimally predicted using a context of at most $240$ ms.



\begin{figure}[htbp]
    \centering
    \subfigure[ACF of composite and filtered EEG and EMG signals. The asterisk indicates the first zero of ACF.]{
        \includegraphics[width=0.46\textwidth]{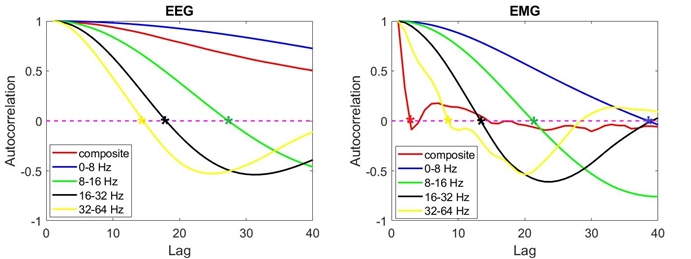}
        \label{figure:acf}
    }
    \subfigure[The values of E1 computed based on the first zero of ACF. The asterisk represents the minimum embedding dimension.]{
        \includegraphics[width=0.46\textwidth]{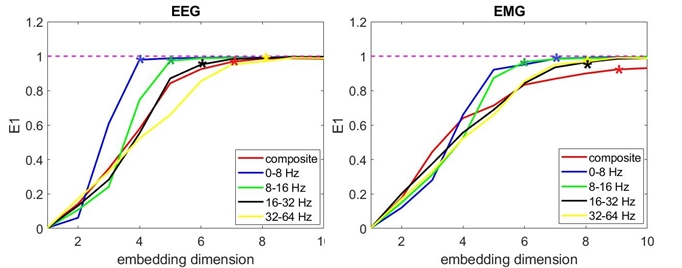}
        \label{figure:cao}
    }
    \caption{\rev{Estimation of the embedding parameters using Cao's criterion. First the embedding delay is estimated as the first zero of the autocorrelation function (top). Given the embedding delay, the embedding dimension is estimated as the dimension at which the E1 metric (bottom) plateaus.}}
\end{figure}

According to Cao's criterion, the embedding delay is set to be the first zero of the autocorrelation function (ACF) of the process, and then the minimum embedding dimension is determined as the dimension beyond which a measure of the relative increase of the  distance between  nearest neighbours remains practically constant \cite{lindner2011trentool,cao1997practical}. Figure \ref{figure:acf} shows the ACFs of the composite and subband EEG and EMG signals from Subject A. For the composite and $(0-8)$ Hz EEG, the first zero of the ACF is outside of the considered range. Since it  was shown in \cite{ragwitz2002markov} that the ACF criterion typically yields $\tau$  that is too large, and moreover according to Ragwitz' criterion $\tau$ is within the investigated range for both signals (see Figure \ref{figure:ragwitz}), we set the embedding delay \(\tau\) of the composite EEG and EEG in the $(0-8)$ Hz range to $\tau=42$. Figure \ref{figure:cao} shows the values of the metric \rev{$E1$} used in Cao's criterion to determine $d$ as the value beyond which $E1$ does not change substantially. It can be seen from the results that for EEG and EMG signals the optimal embedding delay \(\tau\) decreases with increasing frequency, and that an embedding dimension of around $8$ is sufficient to capture the dynamics of the underlying processes. Results obtained using Cao's criterion agree with time-frequency localisation of transient  events and large scale trends. However, the embedding parameters obtained using the two criteria are in stark contrast with each other, often resulting in an order of magnitude difference in the temporal context represented by embedding vectors, and with no clear pattern.

There could be several reasons for these apparent inconsistencies between embedding parameters estimated by the two criteria. Firstly, each criterion, as many others that have been developed for this purpose, makes different specific assumptions about underlying processes, which are difficult to verify in practice, and there is no general consensus on  what the most appropriate criterion would be \cite{pan2020structure}. Secondly, neurophysiological signals are mixtures of processes reflecting different, often unrelated, brain functions, typically degraded by additive noise, which has  a major impact on identifying appropriate embedding parameters for a specific process of interest \cite{ragwitz2002markov}. Finally, the information flow even within a single brain function involves complex interaction within and across frequency bands, including phase-phase, phase-amplitude, amplitude-amplitude, and other forms of coupling \cite{shi2018cross}, that cannot be adequately represented by one embedding vector per frequency band. For instance,  control of human movement depends on a complex interplay between the drive or command from cortical areas and feedback from muscles. Primary motor cortex (M1) feeds forward to spinal motoneurons a command synthesized from other cortical areas such as supplementary motor cortex and prefrontal cortex. This command signal is also fed to the cerebellum which  acts as a comparator or error detector. Signals are fed back from muscle and contain information on muscle length, rate of change of length and tension in muscle attachments. The shortest delay between drive emanating from M1 to, say, a hand muscle is some $20$ ms but other inputs to motoneurons have longer delays. Similarly, information from primary afferents of intrinsic hand muscles signalling muscle length reach sensori-motor cortex at a delay of approximately 20ms, but again slower feedback pathways are also present. Feedback via the cerebellum with varying delays will also be present. Analysis of information flow between cortex, represented by EEG, and muscle, represented by EMG, should therefore ideally take account of bidirectional flow and multiple pathways and delays. 
Therefore, instead of selecting a single set of embedding parameters, we propose TE in the wavelet transform domain by using more than one set of embedding parameters that represent subband processes across different time scales. We will see in Section \ref{sec:teresults} that TE at different time scales reveals complementary information flows.

The proposed approach to multiscale embedding  is to set embedding delays across the wavelet transform scales as
\begin{equation}
\tau_j= s  \tau_j^{(0)},
\label{eq:multiscale}
\end{equation}
where $j$ is the scale index of the wavelet transform, $\tau_j^{(0)}$ is the minimal embedding delay at scale $j$, and $s$ is a positive integer that controls the time scale of the TE analysis. We further propose that the minimal embedding delays across the scales are related according to
\begin{equation}
\tau^{(0)}_{j+1}=2\tau_{j}^{(0)},
\label{eq:twoscale}
\end{equation}
in agreement with the bandwidth of the subband components. \rev{In particular, as the bandwidth of the SWT components increases from low to high-frequency bands, the minimal embedding delays decrease following the same ratio.}
\rev{This two-scale relationship between embedding delays in (\ref{eq:twoscale}) is also in agreement with sampling ratios that would in the case of orthogonal filter banks, or more generally tight frame filters  \cite{daubechies,cvetkovic1998oversampled}, have all the information in neurophysiological signals represented equally by the embedding vectors.} 
An additional degree of freedom towards setting  the time scale, {\sl i.e.} the  temporal context, in TE analysis is the embedding dimension $d$. Whilst for each subband component  a different embedding dimension $d_j$ can be used, we propose using the same dimension $d_j=d$ across all subband components. Since  probability distributions needed for the computation of transfer entropies are estimated from the same data, the same embedding dimension across all subband components ensures that all probability distributions are estimated with the same accuracy. Another aspect of using the same embedding dimension across subband components is that due to the dyadic \rev{relationship between embedding delays } in (\ref{eq:twoscale}) the corresponding embedding captures simultaneously transient events at high frequencies along with larger time scale processes at lower frequencies.

In neurophysiology, interaction delays \(u\) in (\ref{eq:tearbitrarydelayembedding}) can amount to several tens of milliseconds. This is of particular importance for coordinated neural activity and may also affect the results of TE. For the EEG-EMG coupling analysis, the interaction delay is estimated to be between 20 and 35 ms \cite{gao2018electroencephalogram}. Thus, in this study the interaction delay \(u\) was set to correspond to the number of samples in 25 ms for both directions.






Thus, the delay embedding vectors (\ref{eq:delayembeddingvector}) become

\begin{subequations}
\begin{equation}
\mathbf{x}_{j_x,t-1}^{\tau_{j_x},d_x}=(x_{j_x,t-(d_{x}-1) \tau_{j_x}-1}, \dots, x_{j_x,t-1}),
\end{equation}
\begin{equation}
\mathbf{y}_{j_y,t+u-1}^{\tau_{j_y},d_y}=(y_{j_y,t-(d_{y}-1) \tau_{j_y}+u-1}, \dots, y_{j_y,t+u-1}),
\end{equation}
\label{eq:delayembeddingvectormultiscale}%
\end{subequations} 
\rev{where} $x_{j_x,t}$ denotes the sample at time instant $t$ of the subband component $x_{j_x}$ of $x$. Finally
the TE is redefined as
\begin{equation}
\begin{split}
\mathrm{TE}_{x_{j_x} \rightarrow y_{j_y}}(\tau_{j_x}, d_{x}, \tau_{j_y}, d_{y}, u)
=\mathrm{I}(\mathbf{x}_{j_x,t-1}^{\tau_{j_x},d_{x}} ; y_{j_y,t+u} | \mathbf{y}_{j_y,t+u-1}^{\tau_{j_y},d_{y}}) \\
\label{eq:tearbitrarydelayembeddingmultiscale}
\end{split}
\end{equation}
and calculated for all pairs of subband components of $\{x_t\}$ and $\{y_t\}$ and with different choices of embedding vectors, giving the proposed multiscale wavelet TE.

\rev{The term multiscale TE analysis has been used before, but with a different meaning. In \cite{zhao2018multiscale,chen2018multiscale}, the authors employ a graining pre-processing operator that performs local averaging of considered signals. They propose averaging over windows of different lengths prior to the computation of TE and refer to the methodology as multiscale TE. This graining over different time scales has several  merits, but it could also lead to a reduction of available data and spurious detection of causality \cite{zhao2018multiscale}. Moreover, as the local averaging is inherently a low-pass filtering operation, such multiscale graining leads to a loss of high frequency information.
In \cite{shi2018cross}, on the other hand, the authors use the term multiscale correlation to refer just to cross-frequency coupling of electrophysiological processes in general, rather than some specific methodology.}  


\rev{
\section{Evaluations on Simulated Data}
\label{sec:simulation}
Brain functions involve complex interactions across frequency bands, including amplitude-amplitude, phase-phase, and phase-amplitude coupling \cite{shi2018cross}. Therefore, in this section, the effectiveness of the proposed multiscale wavelet TE is evaluated using simulated data generated according to amplitude modulation (AM) and phase modulation (PM) models. Additionally, we compare our method against the MI-in-frequency (MIF) measure \cite{malladi2018mutual}, which is a recent  information-theoretic method for CFC analysis.

\subsection{Amplitude Modulation Model}
\label{sec:am}
Consider amplitude modulation of a carrier signal 
\begin{equation}
c(t)=\sum_{i=1}^{6}a_{c,i}\cos{(2\pi  f_{c,i} t)},
\end{equation}
by a message signal 
\begin{equation}
m(t)=\sum_{i=1}^{6}a_{m,i}\cos{(2\pi  f_{m,i}  t)},
\end{equation}
where the amplitudes \(a_{m,i}\) and \(a_{c,i}\) are generated according to the Rayleigh distribution with the scale parameter $0.5$, whereas the frequencies \(f_{m,i}\) and \(f_{c,i}\) 
are uniformly distributed in the $(1-8)$ Hz and $(33-64)$ Hz frequency ranges, respectively. 
The amplitude modulated carrier signal $s(t)$ is  obtained as
\begin{equation}
s(t)=(1+m(t-u))  c(t),
\end{equation}
\label{eq:am}%
where the interaction delay was set  to $u=9.8$ ms (10 samples at 1024 Hz sampling frequency). 
We evaluated the proposed wavelet transfer entropy  in comparison with the MIF measure by applying them to signals  
\[x(t)=m(t)+w_{x}(t), \quad y(t)=s(t)+w_{y}(t),\]
obtained by degrading 
\(m(t)\) and \(s(t)\) by additive white Gaussian noise sequences  \(w_{x}(t)\) and \(w_{y}(t)\) at different signal-to-noise ratios (SNRs). At each SNR, 10 sets of data were generated, 1s long, and sampled at \(f_{s}=1024\) Hz. Figure \ref{figure:simulation}(a) shows the DFT magnitude of the noise-free message signal \(m(t)\) and amplitude modulated signal \(s(t)\) of a single trial. It can be observed that the spectrum of \(m(t)\) is confined to the  $(0-8)$ Hz band, whereas the spectrum of \(s(t)\) is concentrated predominantly in the $(32-64)$ Hz band. Thus,  we  expect significant TE from the (0-8) Hz band of \(x(t)\) to the (32-64) Hz band of \(y(t)\).

The results of the  wavelet TE and MIF applied to these signals  are shown in Figure \ref{figure:simulation} (c). To allow direct comparison with the MIF technique, the wavelet transfer entropy was calculated for one set of embedding parameters; in particular, the embedding dimension was set to  \(d=6\), and the embedding delays to $\tau=32$, $16$, $8$, and $4$ in $(0-8)$, $(8-16)$, $(16-32)$, and $(32-64)$ Hz bands respectively.  The size of the DFT in the MIF was set as 128  \cite{malladi2018mutual}, to achieve 8 Hz frequency resolution, which is equal to the finest frequency resolution of the wavelet transform. 
It can be observed that the wavelet transfer entropy robustly detects present interactions across all considered noise levels. On the other hand, the ability of the MIF measure to reveal underlying interactions is impaired by the noise, and moreover, some spurious interactions are suggested.

\subsection{Phase Modulation Model}
\label{sec:pm}
To investigate the effectiveness of the proposed method in the context of phase modulation a message signal $m(t)$ is generated as described in the previous subsection, and it is used to generate phase-modulated  signal $s(t)$ according to
\begin{equation}
s(t)=a_{c}\cos(2\pi 50 t+ 3m(t-u)),
\label{eq:pm}%
\end{equation}
where the interaction delay was again set to $9.8$ ms.
The wavelet transfer entropy and MIF were applied to   signals  
\[x(t)=m(t)+w_{x}(t), \quad y(t)=s(t)+w_{y}(t),\]
obtained by degrading 
\(m(t)\) and \(s(t)\) by additive white Gaussian noise sequences  \(w_{x}(t)\) and \(w_{y}(t)\) at different signal-to-noise ratios (SNRs). At each SNR, 10 sets of data were generated, 1s long, and sampled at \(f_{s}=1024\) Hz.
Figure \ref{figure:simulation}(b) shows the DFT magnitude of noise-free message and phase modulated signals of a single trial. The message signal is situated within the $(0-8)$ Hz band, while the spectrum of  phase modulated signal is concentrated in the $(32-64)$ Hz band.

The results of the wavelet transfer entropy and MIF applied to these signals  are shown in Figure \ref{figure:simulation}(d). The parameters of the wavelet TE and MIF are set to the values used in the AM case.  It can be observed from the figure, the proposed TE method robustly detects existing interactions across all noise levels, whereas the MIF measure achieves a partial detection of the interactions and its performance is impaired by the noise.

We considered also MIF with different DFT sizes, {\sl i.e.} different frequency resolutions,  both in the context of amplitude and phase modulation, and did not observe qualitatively different results.

\begin{figure*}[htbp]
    \centering
    \includegraphics[width=0.98\textwidth]{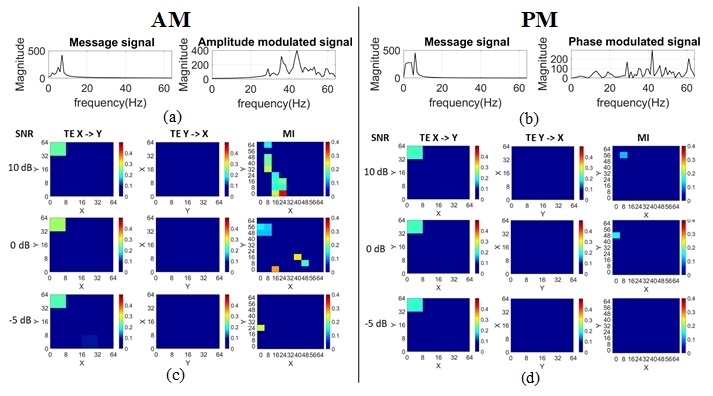}
    \caption{\rev{Evaluation of the wavelet TE and MIF in the context of cross-frequency coupling via amplitude (AM) and phase (PM) modulation. The DFT magnitude of message and modulated signals in AM (a) and in PM (b).  The performance of the wavelet TE and MIF in the AM scenario (c) and PM scenario (d) in the presence of noise. The SNR is set to the same level for both message and modulated signals. For the  TE analysis, the horizontal axis represents the source, whist the vertical axis represents the destination. The colour scale is set so that the dark blue indicates non-significant values.}}
    \label{figure:simulation}
\end{figure*}
}

\section{Evaluations on Neurophysiological Data}
\label{sec:results}
\subsection{Data Acquisition}
\label{sec:data}
The above methodologies were applied to data collected from eight healthy subjects during a motor control task. Standard coherence analysis of these data has been reported previously in \cite{mcclelland2012modulation}.

The subjects gave informed consent for the study, which received ethical approval and was conducted in accordance with the declaration of Helsinki.  All participants were right-hand dominant by self-report. Subjects were seated comfortably at a table and were asked to perform a grasp task with their right hand, holding a 15 cm plastic ruler in a key grip between the thumb and index finger, grasping the end 2 cm of the ruler, and keeping the ruler 2 cm above and parallel to the table surface. Perturbations to the task were generated by an electromechanical tapper, which provided pulses of lateral displacement as described in \cite{mcclelland2012modulation}. 

The length of a single trial was 5 s, with the stimulus given 1.1 s after the start of the trial. The stimuli were generated at pseudorandom intervals varying between 5.6 and 8.4 s (mean 7 s) so that subjects would not predict the arrival of the next stimulus. The entire experiment consisted of up to 8 blocks of 25 trials each. Thus, up to 200 trials of data were collected from each subject. Surface EMG was recorded over the first dorsal interosseous (FDI) muscle of the dominant hand, using self-adhesive electrodes. Bipolar EEG was recorded from the scalp overlying the hand area of the contralateral motor cortex. EEG and EMG signals were sampled at 1024 Hz, amplified and band-pass filtered (0.5-100 Hz for EEG; 5-500 Hz for EMG). Raw data were reviewed offline by visual inspection and epochs of data containing movement or blink artifacts were rejected \cite{mcclelland2012modulation}.

\subsection{CMC and GC Results}
\label{sec:cmcresults}
\label{sec:gcresults}

\begin{figure}[htbp]
    \centering
    \includegraphics[width=0.48\textwidth]{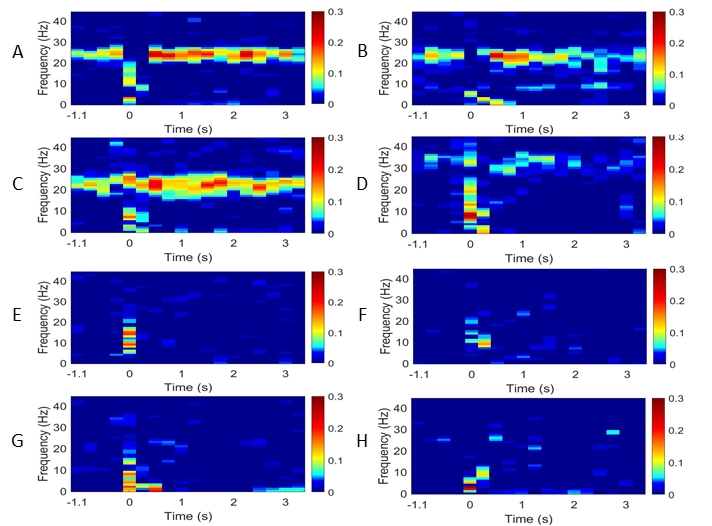}
    \caption{CMC spectrograms in subjects A-H, showing the EEG/EMG coherence. The stimulus is applied time zero. The colour scale is set so that the dark blue  indicates  non-significant coherence values.}
    \label{figure:cmc}
\end{figure}

In the original neurophysiology study \cite{mcclelland2012modulation}, the coherence observed in some subjects was relatively strong, while in others it was very weak. For the present study, eight subjects  that exhibited different levels of coherence were selected.  CMC patterns of the selected subjects are shown in 
Figure \ref{figure:cmc}. The coherence was 
calculated using non-rectified EMG and EEG signals  \cite{Mcclelland2012Rectification,mcclelland2014inconsistent}
and power spectra estimation via the discrete short-time Fourier transform over  \(500\) ms (512 samples) long windows with \(250\) ms (256 samples) overlaps \cite{cvetkovic2000discrete,xu2016corticomuscular}. 
The pronounced increase in CMC across a broad frequency range at the time of the stimulus (time zero in the figures) may in part reflect stimulus artefact and is not considered in this analysis. 
Subjects A-D (top four sub-figures) exhibit significant $\beta$-range  EEG/EMG coherence during the baseline pre-stimulus and post-stimulus period. Subjects E-H, on the other hand, show only a few brief bursts of significant coherence in the early post-stimulus period, and the time-frequency region where CMC is detected is very sparse.

\begin{figure*}[htbp]
    \centering
    \includegraphics[width=0.98\textwidth]{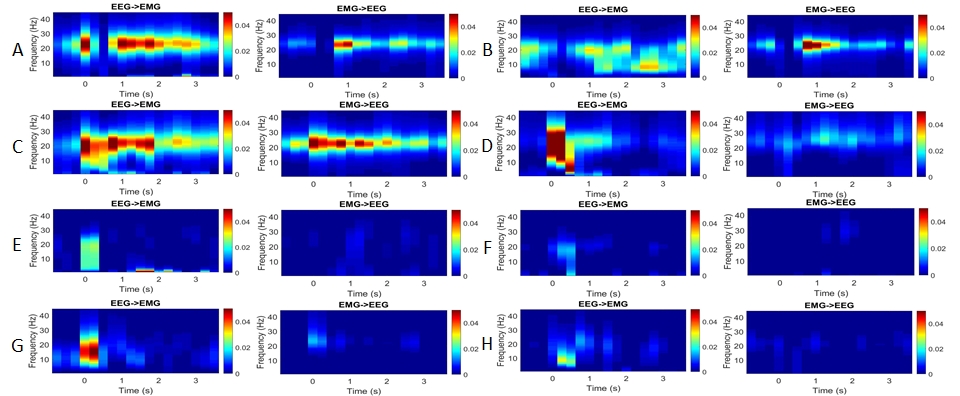}
    \caption{Frequency-domain GC in subjects A-H.  The stimulus was applied at $t=0$. The colour scale is set so that the dark blue indicates non-significant GC.}
    \label{figure:frequencydomaingc}
\end{figure*}

The frequency-domain GC results are shown in Figure \ref{figure:frequencydomaingc}.
\rev{The coefficients of the underlying linear autoregressive moving average (ARMA) models were determined via  least squares model fitting, which is equivalent to 
 maximum likelihood estimation (MLE) assuming Gaussian residuals \cite{barrett2010multivariate,barnett2014mvgc}. Model orders $m$ were determined via the Bayesian Information Criterion (BIC) \cite{schwarz1978estimating} and the Akaike Information Criterion with bias-corrected (AICc) \cite{akaike1974new,hurvich1989regression} which gave $m$ around $30$ for all subjects. The frequency-domain GC was then obtained from the ARMA models following  \cite{geweke1982measurement}.
 All GC analysis was also performed within 500 ms (512 samples) segments every 250 ms (256 samples). As shown in Figure \ref{figure:frequencydomaingc}, 
GC analysis is capable of detecting the direction of information propagation,  however, the overall observation is that results are qualitatively very close to those obtained via the coherence analysis:
%
%
there is a large variability between subjects, making the establishment of some reference interaction patterns very challenging, and more fundamentally, the GC analysis, which like coherence analysis is based on linear models,  is still not sufficiently sensitive to detect the information transfer between cortex and muscle in many healthy subjects.}



\subsection{TE Results}
\label{sec:teresults}
We now investigate the potential of TE for discovering corticomuscular coupling. In order to track temporal evolution of TE, analogous to the CMC and GC analyses, the whole trial (5 seconds) was divided into 18 overlapping segments with time shift of 250 ms (256 samples), each lasting 500 ms (512 samples). Then TE was estimated on each segment separately. \rev{The interaction delay was set to $u=25$ ms, in agreement with the underlying neurophysiology and verified via a delay estimation algorithm based on maximizing TE \cite{wibral2013measuring}.}

\begin{figure*}[htbp]
    \centering
    \subfigure[\rev{TE with the scale factor \(s=4\), which corresponds to the embedding delays of $32$, $16$, $8$, and $4$ samples in the $\delta/\theta$, $\alpha$, $\beta$, and low $\gamma$ bands, respectively.}]{
        \includegraphics[width=0.98\textwidth]{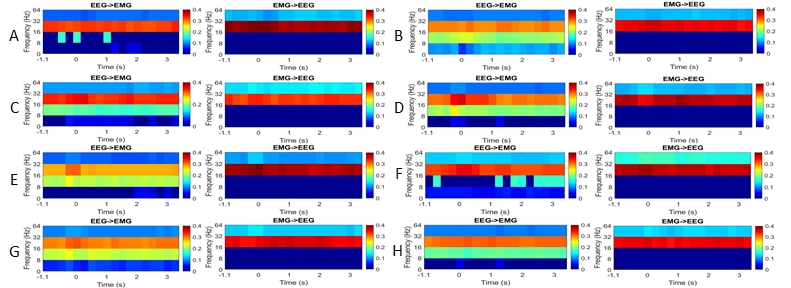}
    }
    \subfigure[\rev{TE with the scale factor \(s=1\), which corresponds to the embedding delays of  $8$, $4$, $2$, and $1$ samples in the $\delta/\theta$, $\alpha$, $\beta$, and low $\gamma$ bands, respectively.}]{
        \includegraphics[width=0.98\textwidth]{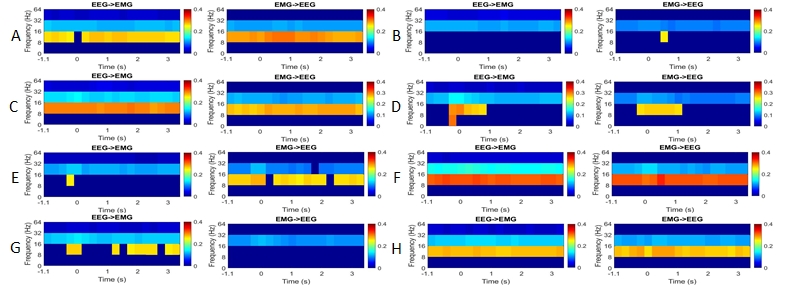}
    }
    \caption{\rev{Multiscale wavelet TE in subjects A-H, showing the EEG \(\rightarrow\) EMG and EMG \(\rightarrow\) EEG TE within  $(32-64)$ Hz - low $\gamma$,  $(16-32)$ Hz  - $\beta$ , $(8-16)$ Hz - $\alpha$, and $(0-8)$ Hz - $\delta/\theta$ frequency bands over time, at two different time scales: $s=4$ (top figure) and $s=1$ (bottom figure). The colour scale is set such that the dark blue indicates non-significant TE values.}}
    \label{figure:frequencydomainte}
\end{figure*}

The multiscale wavelet-domain TE results are shown in Figure \ref{figure:frequencydomainte}.
The wavelet transform was performed at $6$ scales using Daubechies $D4$ filters. Since the signals were recorded using $1024$ Hz sampling, the
subband components  $x_{4,1}$, $x_{5,1}$, and  $x_{6,1}$, approximately 
represent   $(32-64)$ Hz -- low $\gamma$,  $(16-32)$ Hz  -- $\beta$ , and  $(8-16)$ Hz -- $\alpha$ bands, whereas 
$x_{6,0}$  approximately represents the $(0-8)$ Hz band, which are
combined $\theta$ and $\delta$ bands. The decision not to separate $\delta$ and $\theta$ oscillations is governed by data acquisition process during which most of the $(0-4)$ Hz EMG content was filtered out, as is common in motor neurophysiology studies, to minimise movement artefact.
Two different time scale factors \(s=4\) and \(s=1\) were selected for further analysis and the minimum embedding delays $\tau_j^{(0)}$ equal to $8$, $4$, $2$, and $1$ samples in the 
$\delta/\theta$, $\alpha$, $\beta$, and low $\gamma$ bands were used, respectively.
The embedding dimension was set to $d=8$ for all subband components of EEG and EMG signals for both time scales $s=4$ and $s=1$.
To remove the $50$ Hz mains interference, a notch filter was applied. Since the embedding delay differs between frequency bands, the resulting significance level (95\% CL) for individual bands will change accordingly. As shown in the figure, all subjects present similar and significant wavelet-domain multiscale TE throughout the experiment \rev{regardless of whether the subject exhibits a significant coherence/causality}. At the time scale  \(s=4\), the significant TE is mainly concentrated in the $\beta$ range in both directions, but is also  present in the $\alpha$-range from EEG to EMG. With \(s=1\), TE can still be seen in the $\beta$ range, but tends to show stronger coupling in the $\alpha$-range. In particular, $\alpha$-range TE is now also seen in the direction from EMG to EEG, which was not detected with scale factor \(s=4\), illustrating that TE at different time scales reveals complementary interactions.  Fluctuation in the level of TE in relation to the stimulus is observed, particularly in subjects A, D, E, and G.

\begin{figure*}[htbp]
    \centering
    \subfigure[\rev{Cross-frequency TE with the scale factor \(s=4\).}]{
        \includegraphics[width=0.48\textwidth]{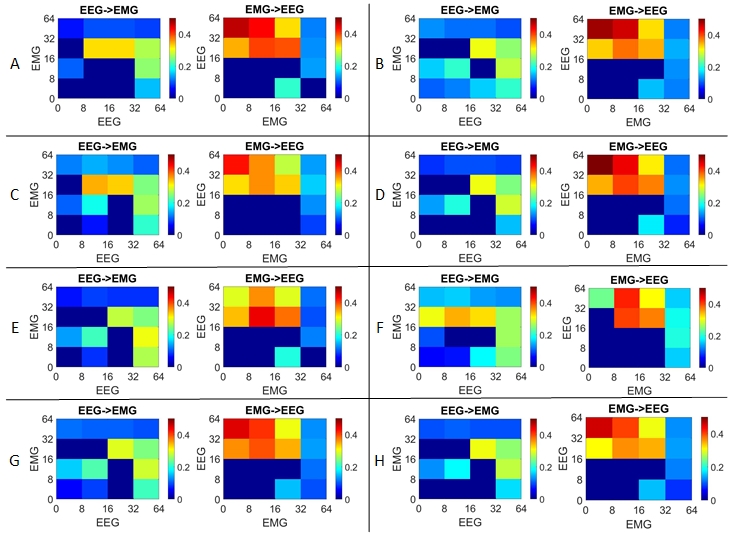}
    }
    \subfigure[\rev{Cross-frequency TE with the scale factor \(s=1\).}]{
        \includegraphics[width=0.48\textwidth]{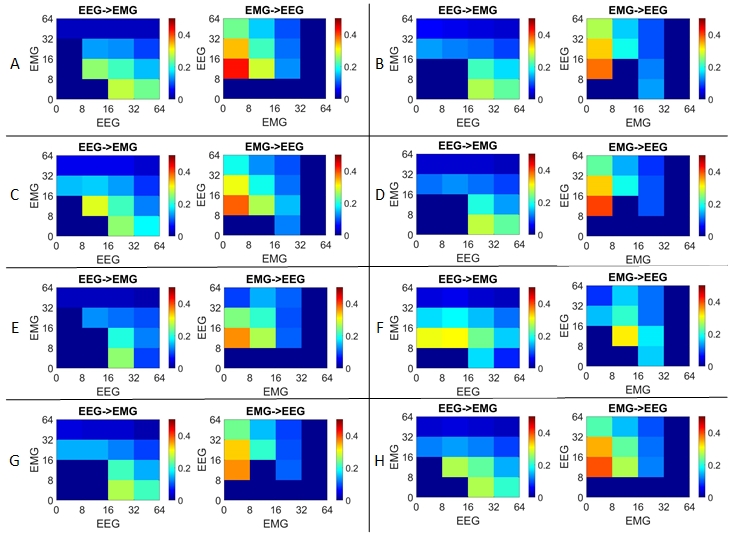}
    }
    \caption{\rev{Multiscale cross-frequency TE in subjects A-H, focusing on four functional bands, $\delta/\theta$, $\alpha$, $\beta$, and low $\gamma$, and the baseline period $[-1.1, -0.6]$ s, at two different time scales: $s=4$ (left figure) and $s=1$ (right figure). The horizontal axis represents the source, whist the vertical axis represents the destination. The colour scale is set so that the dark blue indicates non-significant TE values. }}
    \label{figure:crossfrequencyte}
\end{figure*}

We further investigated \rev{CFC} using the multiscale wavelet TE, focusing on the same four functional bands, $\delta/\theta$, $\alpha$, $\beta$, and low $\gamma$, and the first time segment: the baseline period $[-1.1, -0.6]$ s. 
\rev{The embedding parameters in the CFC analysis were set to the same values as in the intra-band analysis above.} Again, two different temporal scale factors \(s=4\) and \(s=1\) were chosen. The results of this CFC analysis are shown in Figure \ref{figure:crossfrequencyte}. With the scale  \(s=4\), for the direction from EEG to EMG, coupling is mainly seen from EEG-$\beta$ to EMG-$\beta$, and in some subjects also from EEG-$\alpha$ to EMG-$\alpha$ or from EEG-$\alpha$ to EMG-$\beta$; and from EEG-$\gamma$ to EMG-$\alpha$ and $\beta$. In contrast, for the EMG to EEG direction, coupling is observed from EMG-$\delta$/$\theta$ to EEG-$\beta$ and $\gamma$; EMG-$\alpha$ to EEG-$\beta$ and $\gamma$; EMG-$\beta$ to EEG-$\beta$ and $\gamma$. Thus whilst the coupling from EEG to EMG is largely seen within a given frequency band, the coupling from EMG to EEG is seen predominantly across frequency bands. With \(s=1\), for the direction from EEG to EMG, significant TE is mainly concentrated from EEG-$\alpha$ to EMG-$\alpha$ and from EEG-$\beta$ and $\gamma$ to EMG-$\delta$/$\theta$. As for the opposite direction, TE is seen from EMG-$\delta$/$\theta$ maximally to EEG-$\alpha$, and also to EEG-$\beta$, and $\gamma$. Thus there is clear coupling between the lower EMG frequencies and higher EEG frequencies.  
.

\begin{figure}[htbp]
    \centering
    \includegraphics[width=0.46\textwidth]{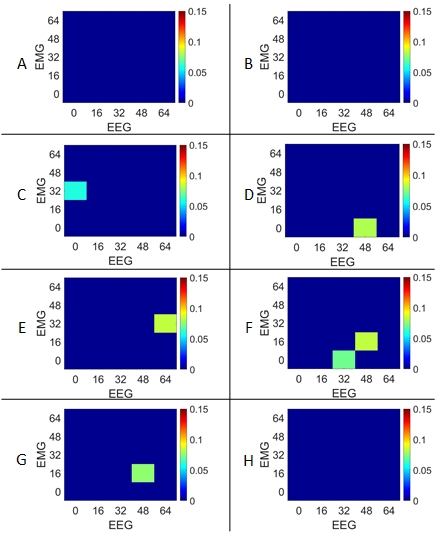}
    \caption{\rev{MIF in subjects A-H during the time interval of [-1.1, -0.6] s. The frequency resolution is 16 Hz. The colour scale is set so that the dark blue indicates non-significant TE values.}}
    \label{figure:mif}
\end{figure}

\rev{For comparison, in  Figure \ref{figure:mif} we show CFC results obtained using the MIF measure on  the same time interval of [-1.1, -0.6] s. The size of the DFT in the MIF analysis is set to 64, corresponding to the frequency resolution of 16 Hz. It can be observed that the MIF  results are inconsistent across subjects, and we cannot observe any significant MI values in subjects A, B, and H.}


\section{Discussion}
\label{sec:discussion}
CMC has previously been shown to vary markedly between subjects \cite{riddle2005manipulation} (also see Figure \ref{figure:cmc}). This inter-subject variability led to the suggestion that coherence was probably measuring linear bidirectional information flows with the relative strength of ascending and descending connections changing between different subjects \cite{witham2011contributions}. GC is a statistical notion of causal influence. In theory, it is entirely equivalent to TE for Gaussian processes and linear couplings \cite{barnett2009granger}. However, because of the neural signals used in this study, the existing directed analysis methods like GC, that are based on linear stochastic models of the data, are still not sensitive enough (also see Figure \ref{figure:frequencydomaingc}). In the present study, the multiscale wavelet TE 
detects cortex-muscle communication in both directions in all subjects tested, including those in whom CMC and GC did not reveal any significant communication. This increased sensitivity to corticomuscular interaction is likely to relate to the ability of TE to detect non-linear as well as linear interactions. This has potential advantages for researchers in motor neuroscience, as $\beta$-range CMC and GC have hitherto been limited by the fact that not all healthy volunteers display significant levels of CMC \cite{salenius2003synchronous,pohja2005reproducibility}. We have previously reported that $\beta$-CMC can be detected following a peripheral stimulus relevant to the motor task, even in those subjects who did not show significant $\beta$-CMC at baseline \cite{mcclelland2012modulation,xu2016corticomuscular}. However, this was often still a small and low amplitude burst of coherence. In contrast, the current study using TE demonstrates significant levels of bidirectional communication between brain and muscle during the baseline and post-stimulus periods, with a consistent pattern between subjects (see Figure \ref{figure:frequencydomainte}). \rev{Although this was a relatively small, exploratory study, and the findings will need to be reproduced in further, larger studies, the findings are potentially of great importance in developing a reliable tool for measuring bidirectional cortex-muscle interactions that could not only transform the field of motor neuroscience, but which could also be translated into clinical practice.}

Considering that neuronal oscillations within each frequency range associate with multiple temporal scales, we proposed using multiple temporal scales to estimate the TE in the wavelet domain, as described in Section \ref{sec:multiscale}. Looking at TE within a given frequency band with \(s=4\), {\sl i.e.} the embedding vectors capture $28$, $56$, $112$ and $224$ ms of temporal context in low $\gamma$, $\beta$, $\alpha$ and $\delta/\theta$ bands, respectively,
TE from EEG to EMG is observed most prominently in the $\beta$ range. Traditional studies using linear methods such as CMC and GC to investigate \rev{cortex-muscle interactions (CMI)} during motor tasks usually find CMC (or GC) in the $\beta$-range \cite{witham2011contributions,riddle2005manipulation,conway1995synchronization,kilner2000human,mcclelland2020abnormal}. This is thought to reflect oscillatory inputs from the motor cortex. However, motor unit firing rates are often lower than this, being around $10-12$ Hz in a modest contraction such as used here, so the presence of a clear EEG to EMG communication also within the $\alpha$ range, as demonstrated here, is not surprising. As for the opposite direction, TE from EMG to EEG appears most prominently in the $\beta$ range, and to a smaller extent in the high-beta/low-gamma range. It is well-established that cortical $\beta$ activity is closely linked with motor control, particularly related to a static task or tonic contraction \cite{brittain2014oscillations,gilbertson2005existing,joundi2012driving} and previous studies of bidirectional \rev{CMI} using GC have demonstrated clear evidence for communication from muscle to EEG in the $\beta$ range \cite{witham2011contributions}.

Interestingly, when a temporal scale  \(s=1\) was applied,  {\sl i.e.} the embedding vectors capture $7$, $14$, $28$ and $56$ ms of temporal context in low $\gamma$, $\beta$, $\alpha$ and $\delta/\theta$ bands, respectively,
a different pattern of communication between EEG and EMG is detected, with TE in both directions being more strongly represented in the $\alpha$-range for most subjects. The reason for this is uncertain, but it is possible that in the context of this shorter timescale, instantaneous drive and feedback on motor unit firing rates is particularly relevant.  

Another important aspect of the current study is the cross-frequency TE results. With \(s=4\), the coupling from EEG to EMG is largely seen within a given frequency band ($\beta$ to $\beta$ and $\alpha$ to $\alpha$), whereas the coupling from EMG to EEG is largely seen across frequency bands ($\delta$/$\theta$ to $\beta$ and $\gamma$; $\alpha$ to $\beta$ and $\gamma$; $\beta$ to $\gamma$) (see Figure \ref{figure:crossfrequencyte}). The observations suggest that CFC is more prominent for communication from EMG to EEG. This could explain in part why CMC and GC (linear methods investigating coupling within the same frequency band) tend to show stronger coupling in the direction from EEG to EMG, rather than in the opposite direction. EMG $\theta$ activity is perhaps more likely to reflect movement or physiological tremor $(5-10)$ Hz, so could carry important information to feedback to the brain reflecting stability of posture. EMG $\alpha$ activity could also reflect this in part, as well as firing of motor units, which tends to range from $10$ to $25$ Hz but can be as low as $6$ Hz and up to $33$ Hz at high levels of contraction. Both these frequency bands from EMG appear to feedback to EEG activity in the $\beta$ range which, as detailed above, is known to relate closely to motor control.

CFC has been extensively studied between different cortical or sub-cortical regions \cite{canolty2010functional}, but is far less widely explored between brain and muscle. The greater CFC seen in the EMG-EEG direction in this study is also in keeping with the non-linear dynamics of sensory systems. In a study of CFC using a modified \rev{CMC} method, Yang {\sl et al.} also found evidence for predominantly linear coupling in the cortex-muscle direction, and of non-linear coupling in the muscle-cortex direction, which they suggested was likely related to sensory feedback pathways \cite{yang2016nonlinear}. Our findings with TE are thus concordant with this. With \(s=1\), clear coupling is observed between the higher EEG frequencies and lower EMG frequencies, which implies that cortex-to-muscle interactions are also characterized to some extent by non-linear activities. This finding is supported by a recent simulation study which revealed both iso-frequency and CFC between supraspinal inputs and motoneurones \cite{sinha2020cross} \rev{and is also concordant with other work demonstrating non-linear activities in the central and peripheral nervous system \cite{stam2003nonlinear,mileusnic2006mathematical,chen2010nonlinear,herrmann2016shaping}.}

The different communication patterns revealed by different temporal scales emphasises the ability of the multiscale TE to detect more complex interaction patterns than is possible using one set of embedding parameters \rev{and which is more in keeping with neuronal models of real brain dynamics \cite{friston2001book}.}

\section{Conclusion}
\label{sec:conclusion}
\rev{In the present study, TE was assessed as a measure of corticomuscular interactions.} In particular,  the concept of multiscale TE in the domain of the dyadic SWT was proposed, that employs multiple sets of embedding parameters to investigate intra- and cross-frequency band coupling at different time-scales. \rev{Experiments using simulated and neurophysiological data substantiate the potential of the multiscale wavelet TE methodology for detecting and quantifying interactions between EEG and EMG processes.
The proposed methodology robustly detected significant information flows for subjects with and without significant CMC or GC. } This can be ascribed to the ability of TE to detect non-linear interactions.
\rev{Results obtained using synchronous EEG and EMG signals are in agreement with the underlying sensorimotor neurophysiology, including the presence of cross-frequency interactions.}
These findings suggest that the concept of multiscale wavelet TE provides a  comprehensive framework for analysing \rev{CMI} and potentially other neurophyisological processes.
\ifCLASSOPTIONcaptionsoff
  \newpage
\fi



%

\bibliographystyle{IEEEtran}

\bibliography{mybibfile}




%








\end{document}